\documentclass[aip,amsmath,amssymb,reprint]{revtex4-1}

\usepackage[version=3]{mhchem} 
\usepackage{xcolor}
\usepackage{caption}
\usepackage{subcaption}
\usepackage{graphicx}
\usepackage{multirow}
\usepackage{amsmath}
\usepackage{pdfpages}

\makeatletter
\AtBeginDocument{\let\LS@rot\@undefined}
\makeatother

\newcommand*\HeHHe[1]{HeHHe$^+$#1}

\newcommand*\kl{k,\lambda}
\newcommand*\SI{Supplementary Material}

\begin{document}
\title{Simulation of Vibronic Strong Coupling and Cavity-Modified Hydrogen Tunneling Dynamics}

\author{Scott M. Garner}
\affiliation{Department of Chemistry, Princeton University, Princeton NJ 08544, United States}
\author{Xiaosong Li}
\affiliation{Department of Chemistry, University of Washington, Seattle WA 98195, United States}
\author{Sharon Hammes-Schiffer}
\email{shs566@princeton.edu}
\affiliation{Department of Chemistry, Princeton University, Princeton NJ 08544, United States}

\begin{abstract}
Polaritons have gained significant attention for the tantalizing possibility of modifying chemical properties and dynamics by coupling molecules to resonant cavity modes to create hybrid light-matter quantum states. 
Herein, we implement the semiclassical nuclear-electronic orbital time-dependent configuration interaction (NEO-TDCI) approach, which treats electrons and specified nuclei on the same quantum mechanical level, while treating the cavity mode classically. 
This \textit{ab initio} dynamics approach can describe both the electronic strong coupling (ESC) and the vibrational strong coupling (VSC) regimes at the same level of theory without invoking the Born-Oppenheimer separation between the quantum nuclei and the electrons.
This approach is used to simulate resonant and off-resonant vibronic strong coupling, where the cavity mode couples to one or many vibronic transitions associated with joint electronic-nuclear excitations within a vibronic progression. 
In this case, the cavity mode couples to nuclear motions even for cavity frequencies typically associated with ESC. 
This approach is also used to illustrate that coupling a molecule to a cavity mode can alter hydrogen tunneling dynamics. 
The semiclassical NEO-TDCI approach provides the foundation for investigating how polaritons may be able to influence chemical reactions involving tunneling and nonadiabatic effects.
\end{abstract}

\maketitle

\section{Introduction}

\begin{figure}[hptb]
    \centering
    \includegraphics[width=\linewidth]{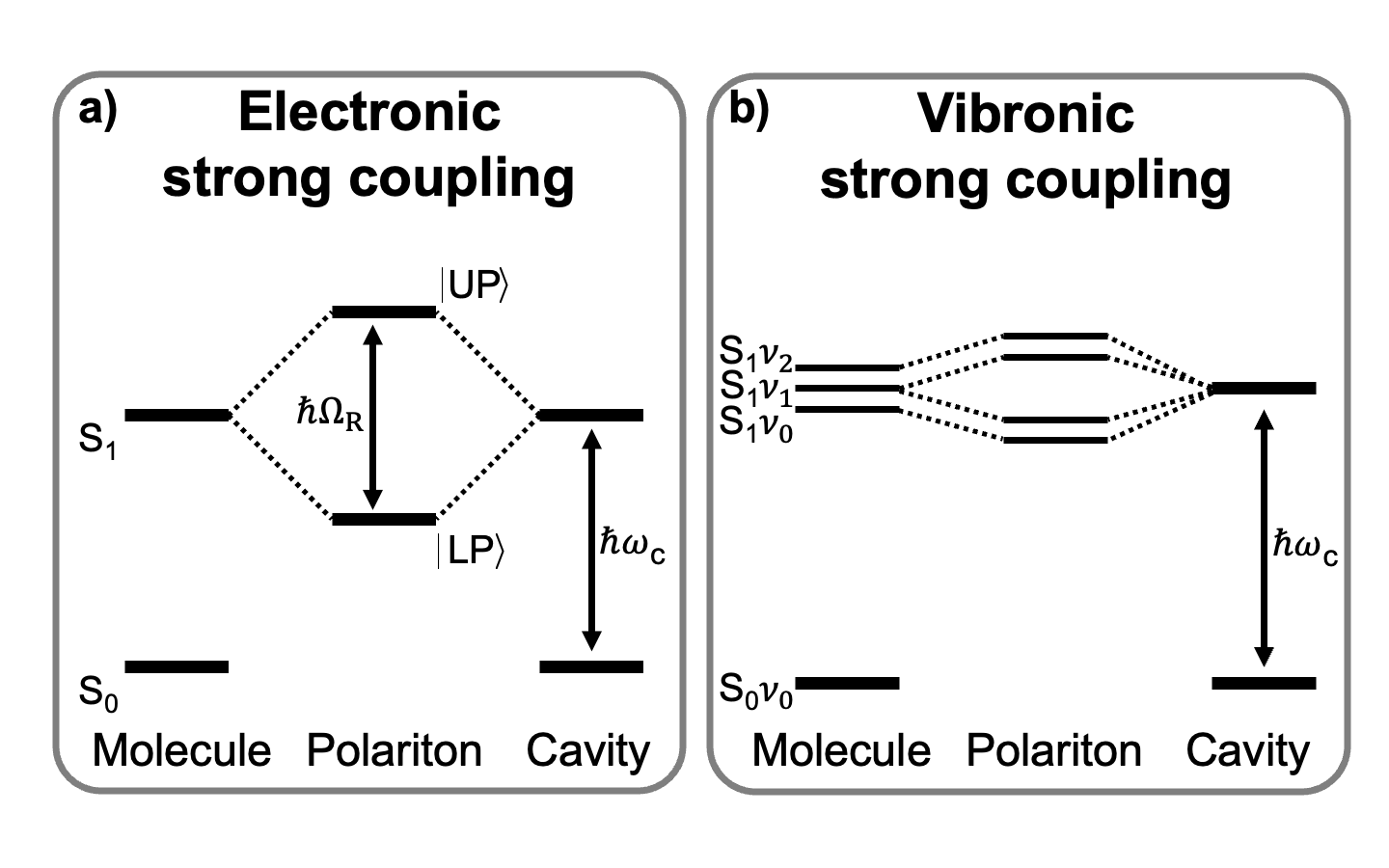}
    \caption{Schematic representation of electronic and vibronic strong coupling.   
    (a) In ESC, the photon is resonant with an electronic transition,  generating two new electronic polaritonic states,  $\left|\text{LP}\right>$ and $\left|\text{UP}\right>$.  
    (b) In vibronic strong coupling, the photon couples to a manifold of vibronic states, generating more than two new vibronic polaritonic states.}
    \label{fig:vbscSchematic}
\end{figure}

Coupling of molecules to a confined electromagnetic field produces hybrid light-matter states known as polaritons\cite{Basov2020_Polariton}.
Polaritons can be formed with ensembles of molecules in optical cavities,\cite{Vahala2003_Optical} where the cavity length is manipulated to tune the resonant frequency to match that of a molecular transition\cite{Skolnick1998_Strong,Raimond2001_Manipulating,Vahala2003_Optical,Hugall2018_Plasmonic,Garcia2021_Manipulating}, or in plasmonic nanocavities, where a plasmonic mode can be coupled  to a single molecule\cite{Chikkaraddy2016_Single}.
A simple understanding of polaritons arises from the Jaynes-Cummings model Hamiltonian,\cite{Jaynes1963_Comparison} where one molecular transition couples to a degenerate cavity mode, hybridizing to form two new light-matter eigenstates, denoted the lower and upper polaritons with Rabi splitting $\Omega_{\rm R}$, as shown in Figure \ref{fig:vbscSchematic}a.
Either electronic strong coupling (ESC)\cite{Hutchison2012_Modifying,Garcia2021_Manipulating} or vibrational strong coupling (VSC)\cite{Long2015_Coherent,George2015_Liquid,Simpkins2023_Control} can be achieved, dependent upon the frequency of the cavity mode.
Each of these regimes is of interest both experimentally and theoretically, motivating a vast array of potential applications.\cite{Basov2020_Polariton,Ebbesen2023_Introduction}

Theoretical modeling of polaritonic chemistry requires understanding the interactions among electrons, nuclei, and cavity modes.
Ideally, all of these subsystems are simultaneously treated quantum mechanically, but exact solutions  quickly become computationally  intractable with system size.
Therefore, approximations are required for first-principles studies of realistic systems.
Quantum electrodynamical (QED) reformulations of conventional quantum chemistry methods\cite{Ruggenthaler2011_Time,Ruggenthaler2014_Quantumelectrodynamical,Haugland2021_Intermolecular,Yang2021_Quantumelectrodynamical,McTague2022_NonHermitian,Pavosevic2022_Wavefunction,Ruggenthaler2023_Understanding,Foley2023_Ab,Liebenthal2023_Assessing,Manderna2024_Comparing,Weidman2024_Cavity,Vu2024_Cavity,Weight2024_Diffusion,Castagnola2024_Strong,Weber2025_Phaseless,Castagnola2025_Strong,Tasci2025_Photon} within the Born-Oppenheimer framework invoke the adiabatic approximation for the electron-nucleus and photon-nucleus interactions but capture the electron-photon interaction without this approximation.
The cavity Born-Oppenheimer approximation (CBO),\cite{Flick2017_Atoms,Flick2017_Cavity,Angelico2023_Coupled,Fischer2023_Cavity,Fiechter2024_Understanding} which groups the photon coordinate with the slow nuclei, further simplifies the electronic structure problem by invoking the electron-photon adiabatic approximation.
In either case, the adiabatic separation between electrons and nuclei persists, which can be detrimental for describing cavity-modified chemistry when nonadiabatic effects are important.
Many studies  have aimed to incorporate electronic-nuclear nonadiabatic effects into polaritonic simulations.
From a conventional molecular dynamics perspective, where nuclei move classically on electronic surfaces, the analogous polaritonic surfaces have been used in surface hopping, \cite{Luk2017_Multiscale,Fregoni2018_Manipulating,Groenhof2018_Coherent,Zhang2019_Nonadiabatic,Fregoni2020_Photochemistry,Fregoni2020_Strong,Hu2022_Quasidiabatic,Lee2024_Formulation} Ehrenfest,\cite{Groenhof2019_Tracking,Zhang2019_Nonadiabatic,Tichauer2021_Multiscale,Hu2022_Quasidiabatic} and \textit{ab initio} multiple spawning\cite{Rana2024_Simulating} simulations.
Polaritonic simulations with quantized nuclei have also been investigated,\cite{Kowalewski2016_Cavitya,Vendrell2018_Coherent,Lacombe2019_Exact,Gu2020_Manipulating,Martinez2021_Case,Fiechter2023_How,Sangiogo2024_Exact} although typically these have been limited to model potentials or diatomic molecules.
To describe vibronic strong coupling and hydrogen tunneling, a fully quantum mechanical description of the nuclei is required, ideally with an even-handed treatment of the electronic and nuclear degrees of freedom.

A powerful approach for incorporating nuclear quantum effects and non-Born-Oppenheimer effects into polaritonic simulations is the nuclear--electronic orbital (NEO) method\cite{Webb2002_Multiconfigurational,Pavosevic2020_Multicomponent,HammesSchiffer2021_Nuclear}, which treats electrons and specified quantized nuclei on equal footing.
In addition to the electronic excitation energies extractable from conventional electronic structure calculations, NEO calculations provide anharmonic vibrational zero-point energy and vibrational excitation energies for the quantized nuclei.
Coupling real-time NEO time-dependent density functional theory (RT-NEO-TDDFT) to a classical cavity mode\cite{Li2022_Semiclassical} produces a unified semiclassical treatment, where both ESC and VSC can be studied at the same level of theory.
This semiclassical methodology captures nonadiabatic interactions between electrons and quantized nuclei, while invoking the electron-photon and quantized nucleus-photon adiabatic approximation. 
Recently, a full-quantum RT-NEO-TDDFT approach, in which the cavity mode is treated quantum mechanically with propagation of a joint molecule-mode density matrix, has also been developed and applied under both ESC and VSC.\cite{Welman2025_LightMatter}

\begin{figure}[hptb]
    \centering
    \includegraphics[width=\linewidth]{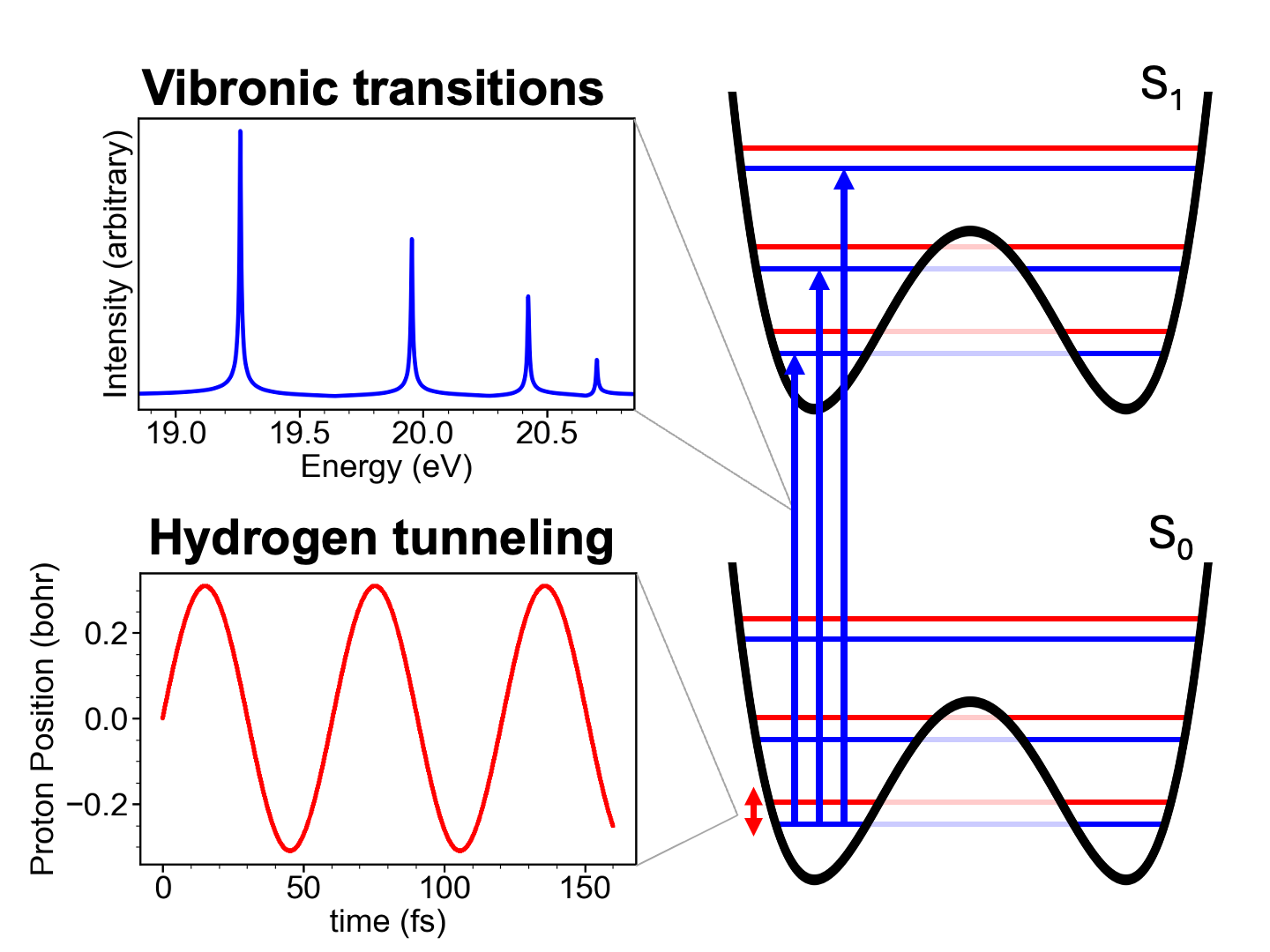}
    \caption{
    Schematic representation of vibronic excitations and hydrogen tunneling for symmetric double-well potential energy curves.
    The vibronic progression in the top spectrum consists of peaks associated with vibronic transitions from the ground proton vibrational state on the ground electronic state to one of the proton vibrational states on the excited electronic state (blue arrows and blue peaks).  
    Only transitions to proton vibrational states with even quantum numbers have non-negligible oscillator strength. 
    The time-dependent hydrogen tunneling signal, corresponding to the expectation value of the proton position operator as a function of time (red oscillatory curve), arises from an initial perturbation at an energy corresponding to the tunneling splitting between the ground and first excited proton vibrational states on the ground electronic state (red double-headed arrow).
    }
    \label{fig:doublewellSchematic}
\end{figure}

In this work, we couple the recently developed NEO time-dependent configuration interaction (NEO-TDCI) method\cite{Garner2024_Nuclear,Garner2025_Timeresolved} to a classical cavity mode to study polaritonic chemistry.
The NEO-TDCI approach is able to capture vibronic progressions due to joint electron-proton excitations and to simulate hydrogen tunneling dynamics\cite{Garner2024_Nuclear}, as depicted in Figure \ref{fig:doublewellSchematic}. 
Both of these phenomena are beyond the theoretical limitations of RT–NEO-TDDFT and linear-response NEO-TDDFT.
Coupling a harmonic cavity mode to each of these quantum dynamical processes allows us to investigate vibronic strong coupling in a cavity and cavity-modified hydrogen tunneling, respectively.
Notably, our semiclassical NEO-TDCI treatment provides a consistent theoretical treatment at these vastly different energy scales, namely eV for vibronic strong coupling versus tens or hundreds of wavenumbers for hydrogen tunneling. 
These energy scales are typically associated with ESC and VSC, respectively. In the NEO framework, however, this distinction is not necessary because electrons and protons are treated quantum mechanically on the same level with no Born-Oppenheimer separation.

In vibronic strong coupling, rather than coupling to a bare electronic transition, the cavity mode is coupled to one or many vibronic transitions associated with joint electronic-nuclear excitations (Figure \ref{fig:vbscSchematic}b).
Vibronic strong coupling has been observed experimentally\cite{Holmes2004_Strong,KenaCohen2008_Strong,Satapathy2021_Selective} but remains difficult to model theoretically, often requiring model Hamiltonians of reduced dimensionality\cite{Fontanesi2009_Organicbased,Mazza2009_Organicbased,Galego2015_CavityInduced,Herrera2017_Dark,Zeb2018_Exact,Ribeiro2018_Polariton} or approximate diabatization schemes.\cite{Vidal2022_Polaritonic}
Our semiclassical NEO-TDCI simulations treat the electronic-nuclear interaction with first-principles quantum chemistry methods and show that the cavity mode couples to the joint electronic-nuclear wavefunction, even though the cavity mode oscillates at frequencies typically associated with electronic motion.
Coupling cavity modes to tunneling has been experimentally observed for electrons in the solid state\cite{Cristofolini2012_Coupling}, with recent work addressing nuclear tunneling using model potentials.\cite{Fiechter2023_How,Ke2024_Insights,Sitnitsky2025_Schrodinger}
Our simulations show that coupling to a cavity can alter tunneling dynamics, opening the possibility of modifying reaction rates via polaritonic cavities.
Overall, this work exemplifies the capabilities of the semiclassical NEO-TDCI method to describe polaritonic phenomena at energy scales commonly associated with both ESC and VSC beyond the Born-Oppenheimer approximation.

The remainder of the paper is organized as follows.
Section \ref{sec:theory} provides the theoretical foundations for our QED Hamiltonian, the NEO-TDCI method, our real-time equations of motion for light and matter, and a method for mapping the molecular vibronic states to the polaritonic states.
Section \ref{sec:computational} provides the computational details of our simulations.
Section \ref{sec:results:vibronic} provides results exemplifying vibronic strong coupling, where the cavity mode is tuned to be resonant with different peaks within a vibronic progression.
Section \ref{sec:results:tunneling} couples the cavity mode to a molecular hydrogen tunneling system that shows modified tunneling behavior based on the light-matter coupling strength.
Finally, Section \ref{sec:conclusions} provides concluding remarks.

\section{Theory}\label{sec:theory}
\subsection{QED Hamiltonian}\label{sec:theory:QEDHam}
Our QED Hamiltonian is written as 
\begin{equation}
    \hat{H}_\text{QED} = \hat{H}_\text{M} + \hat{H}_\text{F}
\end{equation}
where $\hat{H}_\text{M}$ is the molecular NEO Hamiltonian defined in section \ref{sec:theory:NEOTDCI} and $\hat{H}_\text{F}$ is the field Hamiltonian.
In the length gauge and dipole approximation, the field Hamiltonian is\cite{Flick2017_Atoms}
\begin{equation}\label{eq:QEDPFhamiltonian}
    \hat{H}_\text{F} = \frac{1}{2}\sum_{\kl}\left[\hat{p}_{\kl}^2  + \omega_{\kl}^2\left(\hat{q}_{\kl}+\frac{1}{\omega_{\kl}\sqrt{V\epsilon_0}}\hat{\boldsymbol{\mu}}_{\rm M}\boldsymbol{\cdot}\boldsymbol{\xi}_\lambda\right)^2\right]
\end{equation}
The cavity modes are characterized by a wavevector $\mathbf{k}$ of magnitude $k\equiv|\mathbf{k}|$ and a transverse polarization direction represented by the unit vector $\boldsymbol{\xi}_\lambda$ perpendicular to the cavity propagation such that $\mathbf{k}\cdot\boldsymbol{\xi}_\lambda=0$ (i.e., if $\mathbf{k}$ is oriented along $z$, then $\lambda$ is $x$ or $y$).
The field interacts with the molecular system through the total molecular dipole moment $\hat{\boldsymbol{\mu}}_{\rm M}$, which includes all electronic and nuclear contributions to the dipole moment.
The quantities $\hat{q}_{\kl}$, $\hat{p}_{\kl}$ and $\omega_{\kl}$ are the position operator, momentum operator, and frequency of the cavity mode with wavevector magnitude $k$ and polarization direction $\lambda$.
$V$ is the effective cavity volume and $\epsilon_0$ is the vacuum permittivity.

In this work, we use a semiclassical approach, where photons are treated as classical harmonic cavity modes\cite{Li2022_Semiclassical}.
In this approach, the QED Hamiltonian can be rewritten as
\begin{equation}
    \hat{H} = \hat{H}_{\text{sc}} + \frac{1}{2}\sum_{\kl}\left[\hat{p}_{\kl}^2 + \omega_{\kl}^2\hat{q}_{\kl}^2\right].
\end{equation}
The semiclassical light-matter Hamiltonian is given by
\begin{equation}\label{eq:scLMHamiltonian}
    \hat{H}_\text{sc} = \hat{H}_\text{M} + \sum_{\kl} \epsilon_{\kl}q_{\kl}\hat{\mu}_{\lambda}
\end{equation}
where $\hat{H}_{\rm M}$ is again the molecular NEO Hamiltonian.
The light-matter coupling strength is given by $\epsilon_{\kl}=\omega_{\kl}/\sqrt{V\epsilon_0}$, and $\hat{\mu}_\lambda\equiv\hat{\boldsymbol{\mu}}_{\rm M}\cdot\boldsymbol{\xi}_\lambda$ is the total (electronic plus nuclear) dipole operator in the direction of the unit vector $\boldsymbol{\xi}_\lambda$.
We neglect the self-dipole term (i.e., the term proportional to $\hat{\boldsymbol{\mu}}_{\rm M}^2$ in Eq. \ref{eq:QEDPFhamiltonian}) because we are considering only strong coupling as opposed to ultrastrong coupling, where this term becomes important\cite{Flick2017_Atoms,Mandal2020_Polariton,Feist2020_Macroscopic,Schafer2020_Relevance}.

\subsection{NEO-TDCI}
\label{sec:theory:NEOTDCI}
In this subsection, we briefly summarize the key points of the NEO-TDCI method, which has been described in detail previously\cite{Garner2024_Nuclear}.
A NEO-CI wavefunction has the form\cite{Webb2002_Multiconfigurational,Malbon2025_Nuclear}
\begin{equation}
    \left|\Psi^{(n)}_\text{NEO-CI}\right> = \sum_{i,j}C^{(n)}_{i,j}\left|\phi_i^\text{e}\right>\left|\phi_j^\text{p}\right>
\end{equation}
where $\left|\phi_i^\text{e}\right>$ and $\left|\phi_j^\text{p}\right>$ are individual electronic and protonic Slater determinants, respectively, and $C^{(n)}_{i,j}$ are the expansion coefficients for the $n^\text{th}$ time-independent NEO-CI Hamiltonian eigenstate.
In a time-independent calculation, these expansion coefficients are obtained by diagonalization of the NEO-CI Hamiltonian, which contains the kinetic energy of the electrons and quantized nuclei as well as all Coulomb interactions among the classical and quantum particles.

In NEO-TDCI, the time dependence is described via time-dependent expansion coefficients,
\begin{equation}\label{eq:NEOTDCI}
    \left|\Psi_\text{NEO-TDCI}(t)\right> = \sum_{i,j}C_{i,j}(t)\left|\phi_i^\text{e}\right>\left|\phi_j^\text{p}\right>
\end{equation}
while the individual electronic and protonic Slater determinants are time-independent. 
In practice, typically the orbitals from which the determinants are constructed are fixed to be those determined by a NEO Hartree-Fock calculation.
The wavefunction is evolved via numerical integration of the time-dependent Schr\"{o}dinger equation,
\begin{equation}\label{eq:NEOTDCIprop}
    i\frac{d}{dt}\mathbf{C}(t) = \mathbf{H}(t)\mathbf{C}(t)
\end{equation}
where the operator $\hat{H}(t)$ is composed of the time-independent molecular Hamiltonian and any additional external time-dependent perturbations, and $\mathbf{H}(t)$ and $\mathbf{C}(t)$ are represented in the basis introduced in Eq. \ref{eq:NEOTDCI}.

\subsection{Equations of Motion}
\label{sec:theory:EOM}
The cavity mode is evolved according to the equations of motion for a classical harmonic oscillator:
\begin{subequations}
\begin{align}\label{eq:cavityeom}
\dot{q}_{k,\lambda}=&p_{k,\lambda}\\
\dot{p}_{k,\lambda}=&-\omega_{k,\lambda}^2q_{k,\lambda} - \epsilon_{k,\lambda}\mu_\lambda - \gamma_cp_{k,\lambda}
\end{align}
\end{subequations}
where $\gamma_c$ is an effective cavity loss rate.
The interaction with the molecular system is through the total molecular dipole moment $\mu_\lambda(t)$, which includes contributions from all electrons, quantized nuclei, and classical nuclei.  
The total molecular dipole moment is defined as $\mu_\lambda(t)=\left<\hat{\mu}_\lambda(t)\right>-\left<\hat{\mu}_\lambda(t=0)\right>$ where $\left<\cdots\right>$ denotes the operator expectation value. 
This choice sets the effective molecular dipole moment to zero at $t=0$, which removes the interaction between the cavity mode and the permanent molecular dipole moment, ensuring that a non-displaced initial cavity mode (i.e., $p_{\kl}(t=0)=q_{\kl}(t=0)=0$) remains in its ground state in the absence of an external perturbation.\cite{Li2022_Semiclassical}

The NEO-TDCI wavefunction is numerically propagated under the action of the time-dependent semiclassical Hamiltonian: 
\begin{equation}\label{eq:scNEOTDCIprop}
    \left|\Psi\left(t+\delta t\right)\right> =e^{-i\hat{H}_{\rm sc}\delta t}\left|\Psi(t)\right>
\end{equation}
where $\hat{H}_{\rm sc}$ is the semiclassical Hamiltonian of Eq. \ref{eq:scLMHamiltonian}.
The electric field of the cavity mode with effective strength $\epsilon_{k,\lambda}q_{k,\lambda}$ is coupled to both the electronic and nuclear degrees of freedom through the total molecular dipole moment $\hat{\mu}_\lambda$ during propagation of the NEO-TDCI trajectory.
All other aspects of the NEO-TDCI calculation remain identical to those presented in Section \ref{sec:theory:NEOTDCI}. 
Note that when the coupling strength $\epsilon_{k,\lambda}$ is zero, Eq. \ref{eq:scNEOTDCIprop} corresponds to the numerical propagation of Eq. \ref{eq:NEOTDCIprop}.

\subsection{Identifying Contributions from Molecular and Cavity Components to Vibronic Polaritonic States}
\label{sec:theory:mapping}
For a comprehensive analysis of the vibronic polaritonic states, it is useful to identify the contributions from the molecular and cavity components.
Because the cavity mode is treated classically herein, the quantum mechanical weight of the cavity mode is not available.
Thus, we devise a strategy for extracting the contributions of the classical cavity mode to the polaritonic vibronic states.
This strategy is based on the classical cavity mode displacement, $q_{k,\lambda}(t)$, and the molecular wavefunction defined by the expansion coefficients $C_{i,j}(t)$.
To analyze the field character of transitions within the polaritonic spectrum, we inspect the Fourier transform of the cavity mode displacement, $\mathcal{F}\left[q_{k,\lambda}(t)\right]$.
Peaks within this Fourier transform at energies identical to those in the polaritonic spectrum indicate that those transitions contain field character.

To analyze the molecular component of the polaritonic states, we make use of the cavity-free, time-independent molecular vibronic states.
We project the time-dependent coefficients onto the cavity-free molecular vibronic states $\{n\}$ to obtain the time-dependent projected coefficients
\begin{equation}
    C^{(n)}(t) = \left<\Psi^{(n)}_\text{NEO-CI}\middle|\Psi(t)\right> = \sum_{i,j}{C_{i,j}^{(n)}}^\dagger C_{i,j}(t)
\end{equation}
In NEO-TDCI, in the absence of any external perturbation and assuming $\left<\Psi^{(n)}_\text{NEO-CI}\middle|\Psi(t=0)\right>\neq0$, these projected coefficients measure the evolution of each state's complex phase as it evolves under the action of $e^{-i\hat{H}_\text{M}t}$.
In this case, the Fourier transform of the $n^\text{th}$ projected coefficient, $\mathcal{F}\left[C^{(n)}(t)\right]$, contains a single frequency at that eigenstate's \textit{absolute} energy, $E_n$, from which the excitation energy from the ground state to this state can be determined via referencing $E_n$ to the absolute ground state energy, $E_0$.
When coupled to a cavity, the time-dependent wavefunction evolves under the action of $e^{-i\hat{H}_\text{sc}t}$, as given in Eq. \ref{eq:scNEOTDCIprop}, which impacts both the phase and magnitude of the projected coefficients.
The projected coefficients are time-dependent mappings of the unknown eigenstates of $\hat{H}_\text{sc}$ into the known basis of $\hat{H}_{\rm M}$ eigenstates.
The Fourier transforms of these projected coefficients now contain multiple frequencies, which can be mapped to transitions in the polaritonic spectrum by referencing to the ground state energy.
Because the ground state energy may be modified by coupling to the cavity mode, we determine $E_0$ as the primary peak in the Fourier transform of the $C^{(0)}(t)$ projected coefficient. 
In practice, we find minimal difference between this energy and the cavity-free ground state energy.

\section{Computational Details}\label{sec:computational}

The calculations in this work are performed on the \HeHHe{} molecular cation, which is a prototypical hydrogen tunneling system.
Additional calculations of vibronic strong coupling for the HCN molecule are included in the \SI{.}
We utilize a fixed He--He distance of 2.2$~\text{\AA}$, at which the proton moves on a double-well potential energy surface in the conventional Born-Oppenheimer framework\cite{Skone2005_Nuclear}.
In all calculations, the central hydrogen is quantized while the classical helium atoms remain fixed.
We utilize the 6-31G\cite{Hehre1972_Self} electronic and PB4-D\cite{Yu2020_Development} protonic basis sets, with protonic and associated electronic basis function centers placed 0.82$~\text{\AA}$ from each helium atom along the He--He axis.
Our wavefunction \textit{ansatz} for all calculations is a complete active space configuration interaction (NEO-CASCI) with 4 electrons in 8 orbitals and 1 proton in 18 orbitals (excluding orbitals polarized perpendicular to the He--He axis).
The cavity mode is polarized along the He--He axis.

Spectra are calculated from the Fourier transform of the time-dependent dipole moment according to\cite{Goings2018_Real,Li2020_Real}
\begin{equation}
    \sigma(\omega)\propto \omega\left|\mathcal{F}\left[\mu_z(t)e^{-\tau t}\right]\right|
\end{equation}
where $\mu_z$ is the total (electronic plus nuclear) dipole moment along the He--He axis and $\tau$ is a dampening parameter that gives the transitions finite line width.
For all calculations, we utilize $\tau = 10^{-4}$ a.u.
All Fourier transforms are calculated with Padé approximants\cite{Bruner2016_Accelerated} to increase spectral resolution while only running short trajectories.

Our real-time simulations are initialized in three different ways.
To calculate the molecular absorption spectrum outside a cavity, we apply a delta-pulse perturbation with field strength 0.001 a.u. along the He--He axis to the NEO-CASCI ground state.
To calculate spectra inside a cavity, the classical cavity mode is excited by setting $q(t=0)=0.001$ a.u., while the molecular system is initialized in the NEO-CASCI ground state.
Finally, to simulate tunneling, we impart energy into the molecular system by constructing an initial wavefunction superposition of NEO-CASCI vibronic states localized near one of the He atoms, while the cavity begins with no initial displacement (i.e., $p(t=0)=q(t=0)=0$).

The classical cavity mode is propagated via the velocity Verlet algorithm.
No cavity loss is included in the real-time simulations (i.e., $\gamma_c=0$).
The NEO-TDCI propagation is carried out in the basis of determinants via a symplectic split operator integrator.\cite{Blanes2006_Symplectic,Peng2018_Simulating}
For studying vibronic coupling, we utilize a timestep of $\Delta t = 0.005$ a.u. and propagate trajectories for $\approx$ 48 fs.
For simulating tunneling, we utilize a timestep of $\Delta t = 0.01$ a.u. and propagate trajectories for $\approx$ 300 fs.
All NEO-TDCI calculations are performed in a development version of the Chronus Quantum\cite{WilliamsYoung2020_Chronus} software package.

\section{Results and Discussion}
\subsection{Vibronic Strong Coupling}\label{sec:results:vibronic}

We first couple the cavity mode to subpeaks within the vibronic progression associated with the electronic transition from the ground to the first singlet excited state.
Due to molecular symmetry for this $S_0\rightarrow S_1$ electronic transition, the dipole allowed vibronic transitions from the $S_0\nu_0$ state are those with an even number of quanta in the vibrational mode on the excited electronic surface (i.e., the allowed transitions are $S_0\nu_0\rightarrow S_1\nu_0, S_0\nu_0\rightarrow S_1\nu_2, \dots$). 
The cavity-free spectrum showing this progression is shown in Figure \ref{fig:HeHHeVibronicCouplingGrid}d.

We tune the cavity to be resonant with the first three transitions within this vibronic progression, corresponding to the $S_0\nu_0\rightarrow S_1\nu_0$, $S_0\nu_0\rightarrow S_1\nu_2$, and $S_0\nu_0\rightarrow S_1\nu_4$ transitions, which for succinctness we will refer to as cavity frequencies $\omega_0$, $\omega_2$, and $\omega_4$, respectively.
We use three different light-matter coupling strengths: $\epsilon=0.01,0.02,$ and $0.04$ a.u.
The spectra calculated using these parameters are shown in Figure \ref{fig:HeHHeVibronicCouplingGrid}a, b, and c.
All plots contain dashed vertical lines indicating the energies of the relevant NEO-CASCI time-independent Hamiltonian eigenstates and a solid red line indicating the cavity mode frequency.

\subsubsection{Analysis of Vibronic Progression in a Cavity}

\begin{figure*}[hptb]
    \centering
    \includegraphics[width=\linewidth]{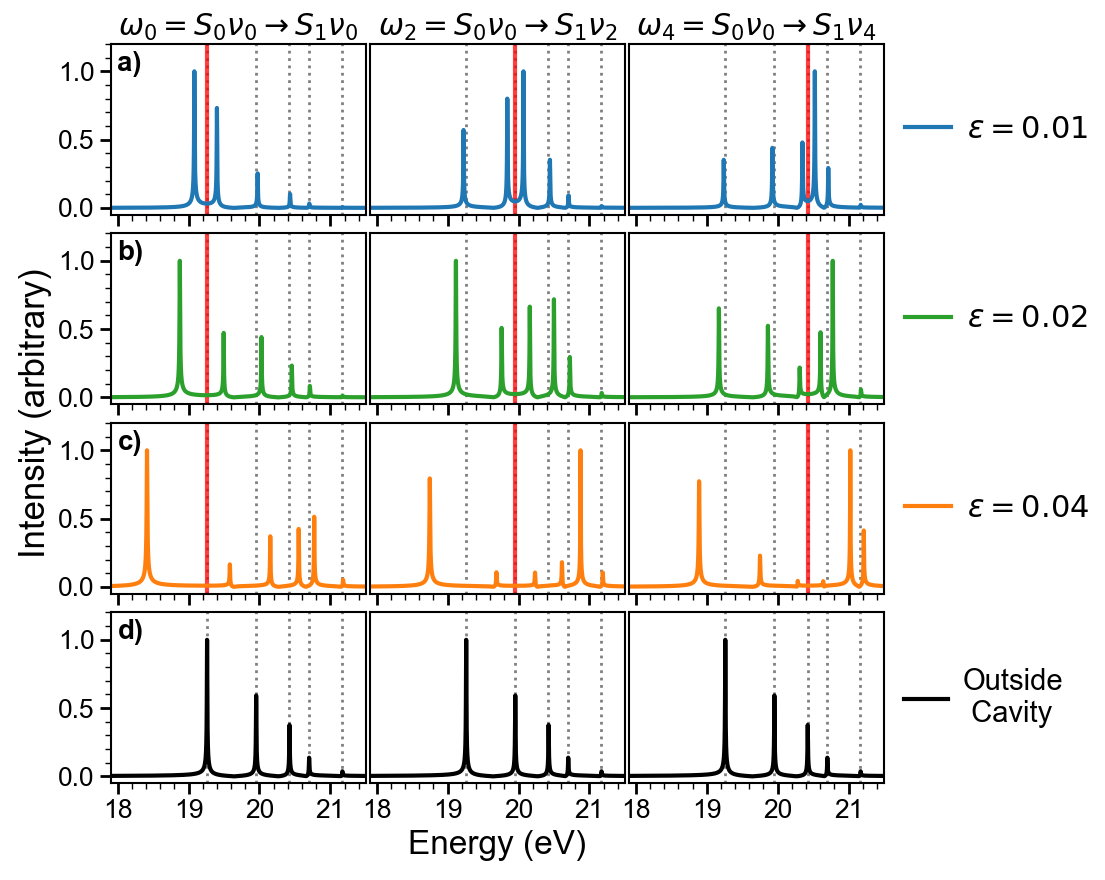}
    \caption{
    Spectra of \HeHHe{} calculated using the semiclassical NEO-TDCI method with three different cavity mode frequencies and three different light-matter coupling strengths, compared to the spectrum outside the cavity.
    The cavity mode frequency is indicated at the top of each column and is shown as a solid red line on the plots.
    The light-matter coupling strength is indicated on the right of each row and increases from (a) 0.01 a.u. to (b) 0.02 a.u. to (c) 0.04 a.u.
    (d) Spectrum of \HeHHe{} calculated outside the cavity.
    All plots in this row are identical, and the dotted vertical lines indicate the energies of the relevant dipole-allowed transitions of \HeHHe{} outside the cavity, as determined from a time-independent NEO-CASCI calculation.
    }
    \label{fig:HeHHeVibronicCouplingGrid}
\end{figure*}

Beginning with the light-matter coupling strength of $\epsilon=0.01$ a.u.  for the three cavity frequencies (Figure \ref{fig:HeHHeVibronicCouplingGrid}a), we observe splitting of the peak corresponding to the vibronic transition to which the cavity is tuned.
The Rabi splittings for the $\omega_0$, $\omega_2$, and $\omega_4$ cavity frequencies are 0.32 eV, 0.23 eV, and 0.17 eV, respectively.
Additionally, in all spectra there are additional peaks of lesser intensity at roughly the molecular transition energies that are not resonant with the cavity mode.
These peaks are always slightly shifted away from the cavity frequency (i.e., if the molecular transition has lower energy than the cavity frequency, the shift is toward lower energy).
We attribute these peaks to the molecular vibronic states that are energetically near the cavity mode but are only very weakly coupled to it, as discussed further below in Section \ref{sec:results:vibronic:mapping}.

At this light-matter coupling, we observe  asymmetry in the Rabi splittings about the cavity frequency.
For $\omega_0$, the lower polariton lies about $0.18$ eV below the cavity frequency, whereas the upper polariton is only $0.14$ eV above the cavity frequency.
These asymmetries can be understood through a perturbation theory perspective, taking the zeroth-order states as the two polaritonic states and the remaining molecular vibronic transitions.
The lower polariton is the lowest energy optically allowed transition and lies in a spectral region free of other transitions, resulting in its peak position being unperturbed by other molecular transitions.
The upper polariton, however, becomes closer to the $S_0\nu_0\rightarrow S_1\nu_2$ transition, which remains primarily molecular in nature, as discussed further in Section \ref{sec:results:vibronic:mapping}.
The interaction between these two states shifts the molecular state to slightly higher energy and the upper polariton to slightly lower energy, leading to the asymmetry in the Rabi splitting.
For $\omega_2$, we observe less asymmetry, with the lower polariton 0.11 eV below and the upper polariton 0.12 eV above the cavity frequency.
In this case, both the upper and lower polaritons have nearby molecular transitions with which they interact to roughly equal degrees.
Finally, for $\omega_4$, the asymmetry is reversed compared to that of $\omega_0$, with the lower polariton 0.07 eV below and the upper polariton 0.10 eV above the cavity frequency.

Moving to the $\epsilon=0.02$ a.u. light-matter coupling strength, the Rabi splitting about the cavity frequency increases for all three cavity frequencies (Figure \ref{fig:HeHHeVibronicCouplingGrid}b).
The intensity and deviation from the cavity-free molecular frequencies for the remaining peaks in the spectra also increase, illustrating off-resonant coupling.
Interestingly, the intensity patterns of the spectra vary for the different cavity mode frequencies.
The spectrum at the $\omega_0$ frequency has a structure resembling that of the cavity-free vibronic progression.
The peaks in the spectrum at the $\omega_2$ frequency do not exhibit a clear pattern, while the peaks in the spectrum at the $\omega_4$ frequency decrease and then increase again on either side of the cavity frequency.

Finally, at the largest light-matter coupling of $\epsilon=0.04$ a.u., we observe more prominent off-resonant coupling, leading to significant alterations to the spectra (Figure \ref{fig:HeHHeVibronicCouplingGrid}c).
The $\omega_0$ cavity frequency spectrum features a clear lower polaritonic peak, but the upper polaritonic peak appears to be at the beginning of a vibronic progression of peaks with increasing energy and intensity.
The $\omega_2$ and $\omega_4$ cavity frequency spectra feature two very intense peaks with many other less intense features.
Interestingly, the two most intense peaks occur at energies below the $S_0\nu_0\rightarrow S_1\nu_0$ transition and above the $S_0\nu_0\rightarrow S_1\nu_6$ transition.
In all spectra at this coupling strength, almost all transitions lie at energies notably different from the cavity-free molecular transition frequencies, with the exception of the $S_0\nu_0\rightarrow S_1\nu_8$ molecular transition at around 21.2 eV.

\subsubsection{Analysis of Molecular and Cavity Components of Vibronic Polaritonic States}\label{sec:results:vibronic:mapping}

\begin{figure*}[hptb]
    \centering
    \includegraphics[width=\linewidth]{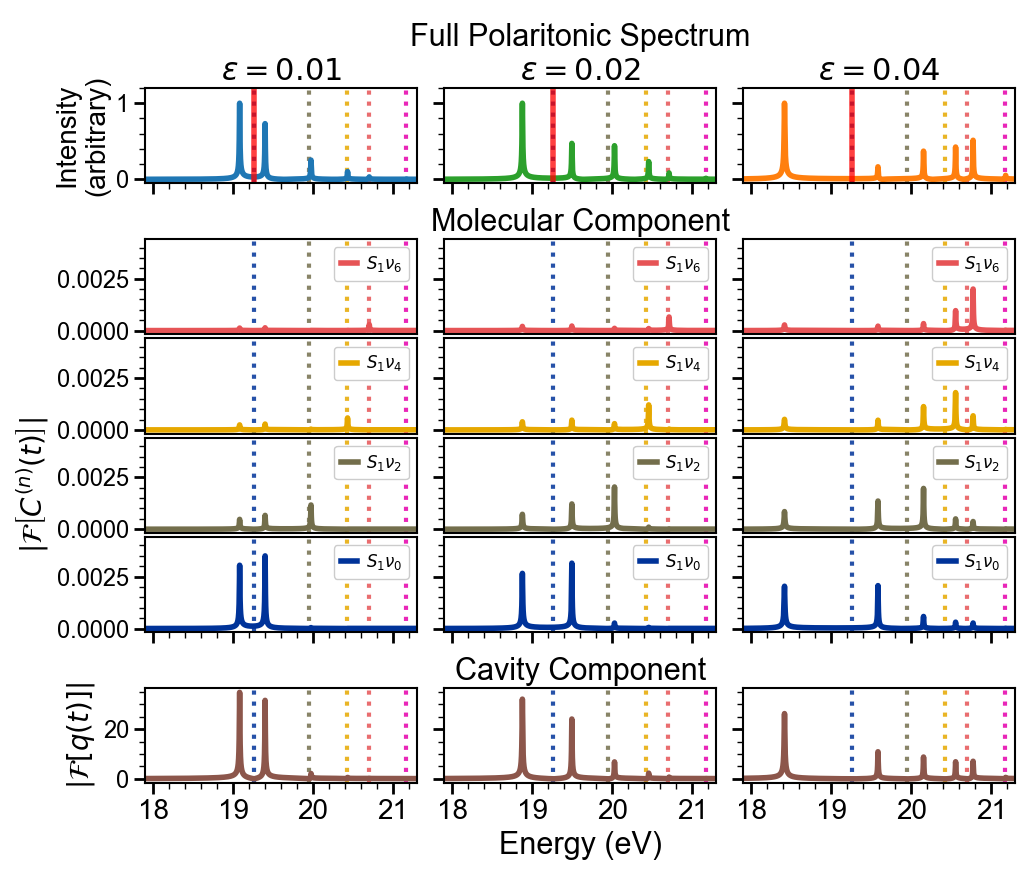}
    \caption{
    Top row: Semiclassical NEO-TDCI spectra of \HeHHe{} calculated with the cavity mode frequency tuned to the $S_0\nu_0\rightarrow S_1\nu_0$ transition (solid red lines) and the light-matter coupling constant indicated above each column.
    Middle set of plots: Fourier transforms of the projected coefficients of the relevant molecular vibronic states for the light-matter coupling constants indicated above each column.
    Bottom row: Fourier transforms of the classical cavity mode displacement coordinate.
    On all plots, the dotted vertical lines indicate the energies of the relevant dipole-allowed transitions of \HeHHe{} outside the cavity, as determined from a time-independent NEO-CASCI calculation, color coded to match the relevant state from the middle panels.
    }
    \label{fig:HeHHeAnalysisOmega1}
\end{figure*}

\begin{figure*}[hptb]
    \centering
    \includegraphics[width=\linewidth]{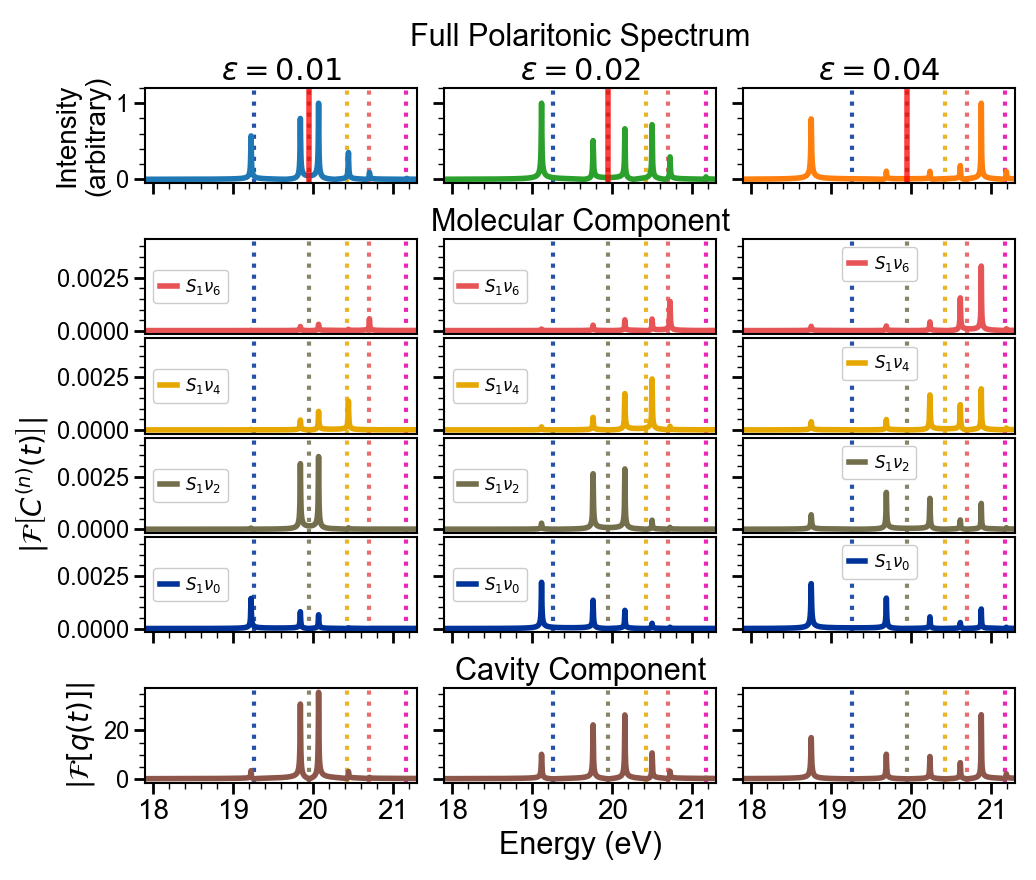}
    \caption{
    Top row: Semiclassical NEO-TDCI spectra of \HeHHe{} calculated with the cavity mode frequency tuned to the $S_0\nu_0\rightarrow S_1\nu_2$ transition (solid red lines) and the light-matter coupling constant indicated above each column.
    Middle set of plots: Fourier transforms of the projected coefficients of the relevant molecular vibronic states for the light-matter coupling constants indicated above each column.
    Bottom row: Fourier transforms of the classical cavity mode displacement coordinate.
    On all plots, the dotted vertical lines indicate the energies of the relevant dipole-allowed transitions of \HeHHe{} outside the cavity, as determined from a time-independent NEO-CASCI calculation, color coded to match the relevant state from the middle panels.
    }
    \label{fig:HeHHeAnalysisOmega3}
\end{figure*}

Using the methodology presented in Section \ref{sec:theory:mapping}, we investigate the relationship between the molecular and polaritonic eigenstates.
The analyses for the $\omega_0$ and $\omega_2$ cavity frequencies are shown in Figures \ref{fig:HeHHeAnalysisOmega1} and \ref{fig:HeHHeAnalysisOmega3}, respectively, and the analysis for the $\omega_4$ cavity frequency is presented in the \SI{.}
For each of these cavity frequencies, we calculate projected coefficients for the relevant time-independent molecular states present in the cavity-free spectrum (i.e., the first four vertical dashed lines shown in the panels of Figure \ref{fig:HeHHeVibronicCouplingGrid}). 
Moreover, a similar analysis of vibronic coupling for the HCN molecule is provided in the \SI{.}

We begin our discussion with the $\omega_0$ cavity frequency.
At the smallest light-matter coupling (Figure \ref{fig:HeHHeAnalysisOmega1}, left column), we observe clear responses at the energies of the polaritonic transitions ($19.08$ and $19.40$ eV) in both the cavity and molecular components, confirming the hybrid light-matter nature of the the polaritonic states.
Based on the relative intensities of the different vibronic state projected coefficients, the primary contribution from the molecular eigenstates arises from the $S_0\nu_0\rightarrow S_1\nu_0$ transition, which is not surprising because the cavity frequency is resonant with this transition. 
The transition to the $S_1\nu_2$ vibronic state, and to a lesser extent the higher lying vibronic transitions, also exhibits minor peaks at the polaritonic energies.
At higher energies, we observe minimal response from the cavity component, implying that the transitions at these energies in the calculated spectrum are primarily molecular in nature, with a small contribution from the cavity component to the peak at $\approx 20$ eV.
This analysis also confirms our interpretation of the asymmetry of the polaritonic peaks relative to the cavity mode energy.  
In particular, although the Rabi splitting is due primarily to coupling between the $S_0\nu_0\rightarrow S_1\nu_0$ vibronic transition and the cavity mode, some mixing of these two transitions with the higher-lying vibronic transitions is still present.
Thus, even at the smallest value of the light-matter coupling used, a simple two-level Jaynes-Cummings type model is insufficient to describe vibronic strong coupling, as clearly many molecular vibronic transitions must be considered for a complete understanding of the polaritonic spectrum.

At the intermediate light-matter coupling strength (Figure \ref{fig:HeHHeAnalysisOmega1}, center column), we begin to observe more peaks in all of the Fourier transforms.
Inspecting the cavity component, in addition to peaks associated with the Rabi splitting at $\approx 18.87$ and 19.50 eV, there are additional peaks in the Fourier transform of the cavity displacement, reflecting off-resonant coupling.
Just above the $S_0\nu_0\rightarrow S_1\nu_2$ transition energy, the cavity response indicates that the corresponding transition in the spectrum has notable cavity mode character.
At the largest light-matter coupling (Figure \ref{fig:HeHHeAnalysisOmega1} right column), all peaks in the calculated NEO-TDCI polaritonic spectrum (top row) have corresponding features in the Fourier transform of the cavity displacement (bottom row), indicating that all of the observed transitions now have polaritonic character.
This finding is further supported by inspecting the molecular components, where all relevant molecular vibronic states exhibit a response at all energies in the NEO-TDCI polaritonic spectrum (top row), indicating that the cavity mode directly couples to all vibronic transitions.

Next, we perform this analysis for coupling the cavity mode to the $S_0\nu_0\rightarrow S_1\nu_2$ transition.
At the smallest value of the light-matter coupling (Figure \ref{fig:HeHHeAnalysisOmega3}, left column), the polaritonic states at 19.84 and 20.07 eV again clearly contain both cavity and molecular components.
The transitions to the $S_1\nu_0$, $S_1\nu_4$, and $S_1\nu_6$ vibronic states appear to be primarily molecular in nature, as the cavity component shows minimal response at these energies, although small peaks at the transitions to the $S_1\nu_0$ and $S_1\nu_4$ vibronic states are evident.
In this case, more molecular states are coupled to the polaritonic states, as the projected coefficients for the  $S_1\nu_0$ and $S_1\nu_4$ show noticeable intensity at the polaritonic energies.
For this transition, we observe the most symmetric Rabi splitting.
In this case, the primary polaritonic peaks couple to both lower ($S_0\nu_0\rightarrow S_1\nu_0$) and higher ($S_0\nu_0\rightarrow S_1\nu_4$) energy molecular vibronic transitions, as these states show approximately equal responses at both polaritonic frequencies.
From a perturbation theory perspective, where the coupling strength is controlled by both a perturbative matrix element numerator and energy difference denominator, the $S_0\nu_0\rightarrow S_1\nu_0$ molecular transition has the larger transition dipole moment (to which the cavity mode couples) while the $S_0\nu_0\rightarrow S_1\nu_4$ transition is energetically closer to the cavity frequency.
These competing factors in this case lead to roughly equal coupling strengths to both the upper and lower polaritons, explaining the symmetry in the splitting.
At the intermediate and largest light-matter coupling strengths (Figure \ref{fig:HeHHeAnalysisOmega3}, center and right columns), the behavior is similar to that for the $\omega_0$ cavity frequency: as the light-matter coupling is increased, more transitions in the spectrum have appreciable cavity mode character, and the peaks in the polaritonic spectrum arise from the mixing of many molecular vibronic states.

\subsection{Hydrogen Tunneling}\label{sec:results:tunneling}

The transfer of a proton from a donor to an acceptor represents a fundamental chemical reaction motif where tunneling effects can be important\cite{Layfield2014_Hydrogen,HammesSchiffer2025_Explaining}.
The NEO-TDCI method has previously been used to simulate the real-time, nonequilibrium dynamics of hydrogen tunneling.\cite{Garner2024_Nuclear}
We now couple these tunneling dynamics to a cavity mode for the \HeHHe{} molecule with a fixed He--He distance.
From the perspective of the conventional Born-Oppenheimer approximation, the tunneling proton moves on a symmetric double-well electronic potential energy surface, and the ground and first excited proton vibrational wavefunctions are symmetric and antisymmetric, respectively, across the two wells (Figure \ref{fig:doublewellSchematic}). 
In the NEO framework, the lowest two vibronic eigenstates correspond to analogous symmetric, $\left|\Psi_{\rm S}\right>$, and antisymmetric, $\left|\Psi_{\rm A}\right>$, nuclear-electronic wavefunctions, where the symmetry applies to the nuclear portion of the wavefunction.  
The  energy difference between these two vibronic states is the tunneling splitting.

In order to simulate tunneling dynamics, we construct an initial molecular wavefunction superposition as
\begin{equation}
    \left|\Psi(t=0)\right> = \frac{1}{\sqrt{2}}\left(\left|\Psi_{\rm S}\right>+i\left|\Psi_{\rm A}\right>\right)
\end{equation}
where the factor of $i$ ensures that the wavefunction at $t=0$ has zero dipole moment and therefore no interaction with a cavity mode.
Because we are imparting energy via our construction of a nonequilibrium initial wavefunction, the cavity is initialized with no momentum or displacement: $p(t=0)=q(t=0)=0$.

\begin{figure}[hptb]
    \centering
    \includegraphics[width=\linewidth]{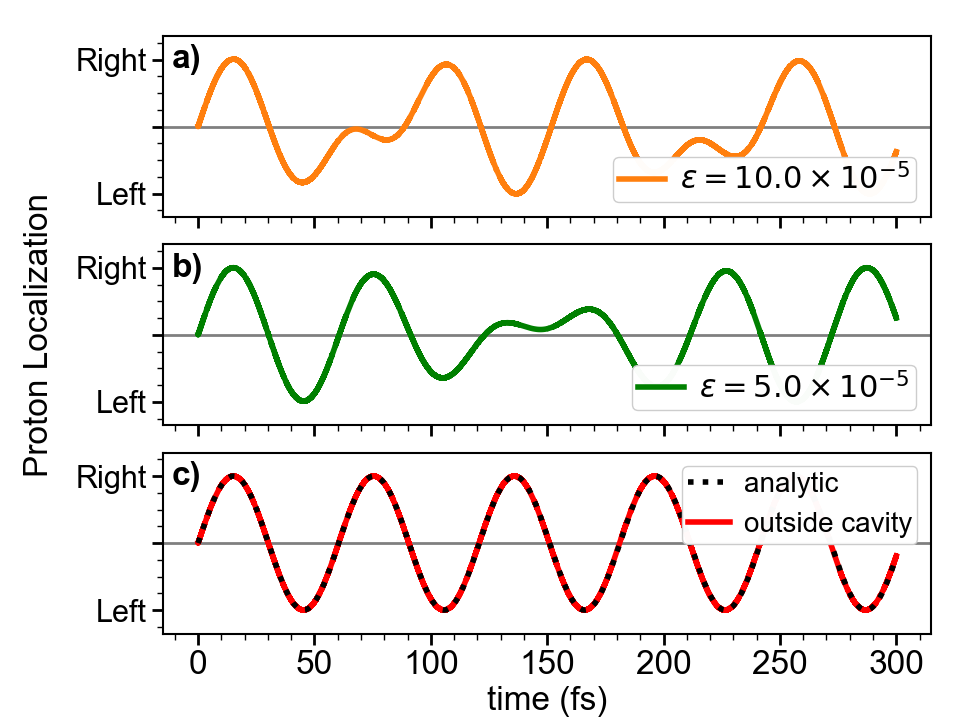}
    \caption{
    (a, b) \HeHHe{} tunneling dynamics when coupled to a classical cavity mode with two different light-matter coupling strengths.
    The given light-matter coupling strengths are in atomic units.
    (c) \HeHHe{} tunneling dynamics outside the cavity, along with the analytic solution for a two-level system. 
    The proton dipole moment is used as a proxy for the proton localization.  
    Note that the proton density is bilobal and delocalized between the two He atoms at the y-axis value midway between the right and left localizations.
    }
    \label{fig:HeHHeTunnelingCoupled}
\end{figure}

Outside a cavity, propagating this initial wavefunction superposition leads to the oscillation of the proton density between the donor and acceptor He atoms. 
We will use the time-dependent proton dipole moment as a metric for proton localization.
Analytically, it can be shown\cite{Garner2024_Nuclear} that for this two-level system, the time-dependent proton dipole moment should evolve according to $D\sin\left(\Delta Et\right)$, where $\Delta E$ is the tunneling splitting and $|D|$  is the maximum magnitude of the proton dipole moment, which occurs when the proton is localized near one of the He atoms. 
The time-dependent proton dipole moment of a cavity-free NEO-TDCI simulation, as well as the analytic solution, is shown in Figure \ref{fig:HeHHeTunnelingCoupled}c.

We now couple the \HeHHe{} to a classical cavity mode tuned to the energy of the tunneling splitting, which is $\approx 552$ $\text{cm}^{-1}$ for the geometry, basis set, and active space used.
Because the proton tunneling creates a large change in dipole moment, significantly smaller light-matter couplings than those used in the vibronic strong coupling case are required to observe cavity effects on the tunneling dynamics.
The time-dependent proton dipole moments for two different coupling strengths are shown in Figures \ref{fig:HeHHeTunnelingCoupled}a and b.
For both coupling strengths, the initial $\approx 50$ fs of the trajectory is very similar to that outside the cavity.
After this time, however, the trajectories coupled to a cavity mode display different behavior.
For both light-matter coupling strengths, the oscillations are dampened between $\approx50-100$ fs for $\epsilon=10.0\times10^{-5}$ a.u. and $\approx100-175$ for $\epsilon=5.0\times10^{-5}$ a.u.
After these periods of time, the dipole returns to oscillating at the same frequency as outside the cavity but with opposite phase (i.e., localized on the opposite side as the cavity-free case).
Between $\approx 180-240$ fs in the $\epsilon=10.0\times10^{-5}$ a.u. trajectory, the oscillations are dampened again, after which the proton returns to oscillating with the same phase as that of the cavity-free trajectory.

\begin{figure*}[hptb]
    \centering
    \includegraphics[width=0.8\linewidth]{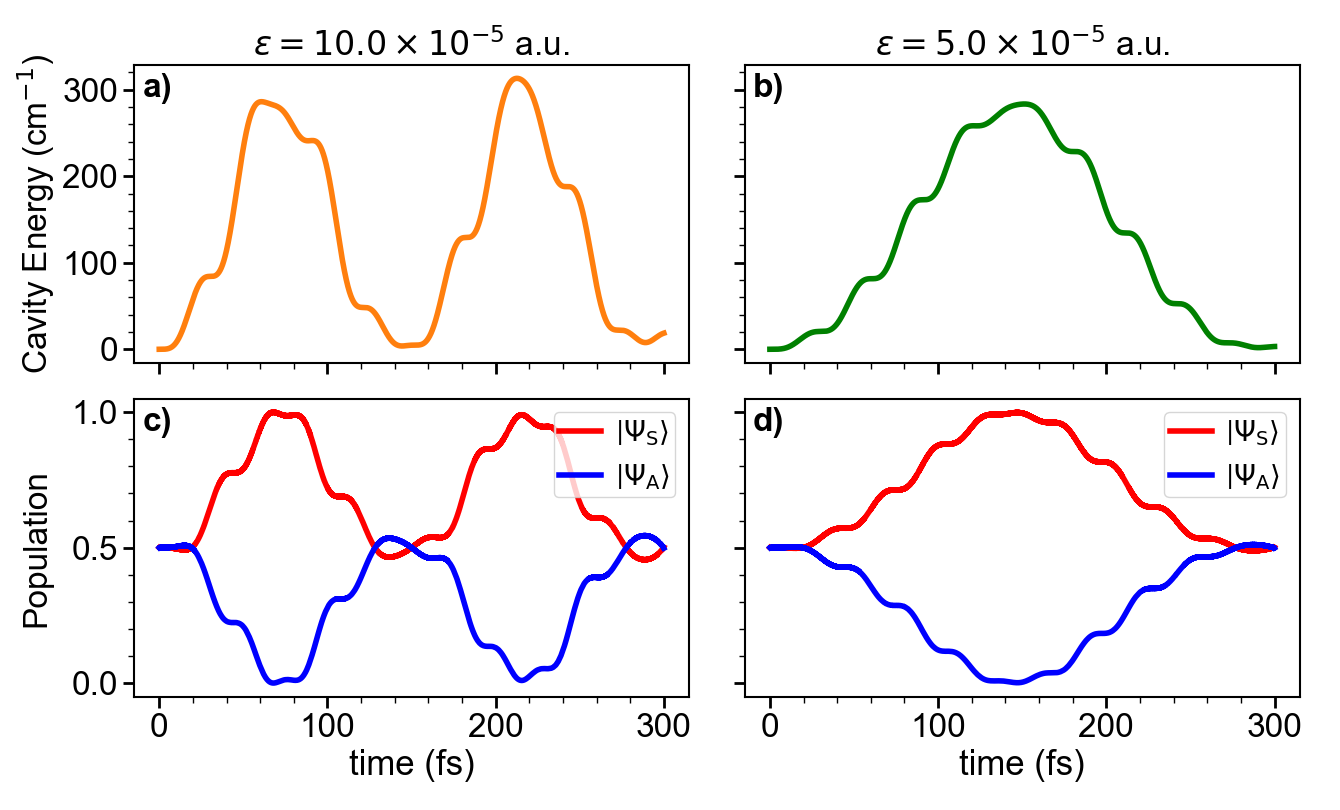}
    \caption{
    (a,b) Cavity mode energy and (c,d) populations of $\left|\Psi_{\rm S}\right>$ and $\left|\Psi_{\rm A}\right>$ for the cavity-coupled \HeHHe{} tunneling trajectories shown in Figures \ref{fig:HeHHeTunnelingCoupled}a and b.
    }
    \label{fig:HeHHeTunnelingCavityDisplacement}
\end{figure*}

For both light-matter coupling strengths, the dampening of the oscillations is attributable to the transfer of energy from the molecular excitation to the cavity mode.
Figures \ref{fig:HeHHeTunnelingCavityDisplacement}a and b show the cavity mode energy, including the kinetic and potential energies, for these trajectories.
Inspecting the populations of the $\left|\Psi_{\rm S}\right>$ and $\left|\Psi_{\rm A}\right>$ states in the time-dependent wavefunction (Figure \ref{fig:HeHHeTunnelingCavityDisplacement}c and d), we can confirm that energy has transferred between the molecule and the cavity mode, as the population resides primarily in the lower-energy $\left|\Psi_{\rm S}\right>$ state when the cavity is at its maximum energy.

These trajectories exemplify how vibrational strong coupling can influence reaction dynamics.
The imparting of energy via the inclusion of the vibrationally excited state $\left|\Psi_{\rm A}\right>$ in the initial wavefunction leads to the tunneling dynamics, and coupling the cavity mode to this tunneling frequency leads to an efficient exchange of energy between the molecular vibrations and the cavity mode.
Although in this case the cavity mode  returns the energy to the molecular system, which then continues to oscillate, the presence of  additional vibrational degrees of freedom, such as modes that alter the donor-acceptor distance for a proton transfer reaction, could lead to either suppression or enhancement of the proton transfer reaction.

\section{Conclusions}\label{sec:conclusions}

We implemented the semiclassical NEO-TDCI approach and used it to investigate the coupling of a cavity mode to vibronic progressions and hydrogen tunneling dynamics.
This approach quantizes select nuclei at the same level as the electrons without invoking the Born-Oppenheimer approximation between them, while treating the cavity mode classically.
For the vibronic progression studied herein, we observe direct coupling between the cavity mode and individual vibronic transitions, leading to typical Rabi splittings between the polaritonic peaks at small coupling strengths. 
At larger coupling strengths, we observe off-resonant coupling of the cavity mode to many vibronic transitions simultaneously, leading to more complex polaritonic spectra.
These observations implicate coupling of the cavity mode to nuclear motions even for cavity frequencies typically associated with ESC.
These simulations also indicate that inclusion of nonadiabatic interactions between electrons and nuclei is critical for appropriately describing vibronic strong coupling.

Additionally, we showed how coupling to a cavity mode can impact the dynamics of hydrogen tunneling, with different coupling strengths leading to varying proton dynamics.
The NEO multireference configuration interaction approach\cite{Malbon2025_Nuclear,Stein2025_Computing} has recently been shown to be able to accurately predict tunneling splittings, providing an avenue for quantitatively studying cavity-modified chemical reactivity when tunneling is a key mechanistic step. Overall, this work underscores the need for a consistent quantum mechanical treatment of electrons and nuclei that extends beyond the Born-Oppenheimer approximation in order to understand how polaritons can fundamentally alter chemical properties.

\section{Acknowledgement}
This application to polaritons is based upon work supported by the Air Force Office of Scientific Research under AFOSR Award No. FA9550-24-1-0347.
The development of NEO-TDCI methods in the Chronus Quantum computational software was supported by the Department of Energy in the Computational Chemical Science program (Grant No. DE-SC0024935). 
The software infrastructure, including NEO integrals and self-consistent-field, was supported by the Office of Advanced Cyberinfrastructure, National Science Foundation (Grant Nos. OAC-2103717, OAC-2103902, and OAC-2401207). 
We thank Jonathan Fetherolf and Millan Welman for helpful discussions.

\section{Supplementary Material}
The \SI{} contains additional analysis of the \HeHHe{} polaritonic spectra for the $\omega_4$ cavity frequency and calculations of vibronic strong coupling spectra and associated analysis for the HCN molecule.

\section{Data Availability}
The data that support the findings of this study will be openly available on Zenodo upon publication.

\nocite{*}
\bibliography{scNEOTDCI}
\includepdf[pages={{},1,{},2,{},3,{},4,{},5,{},6,{},7}]{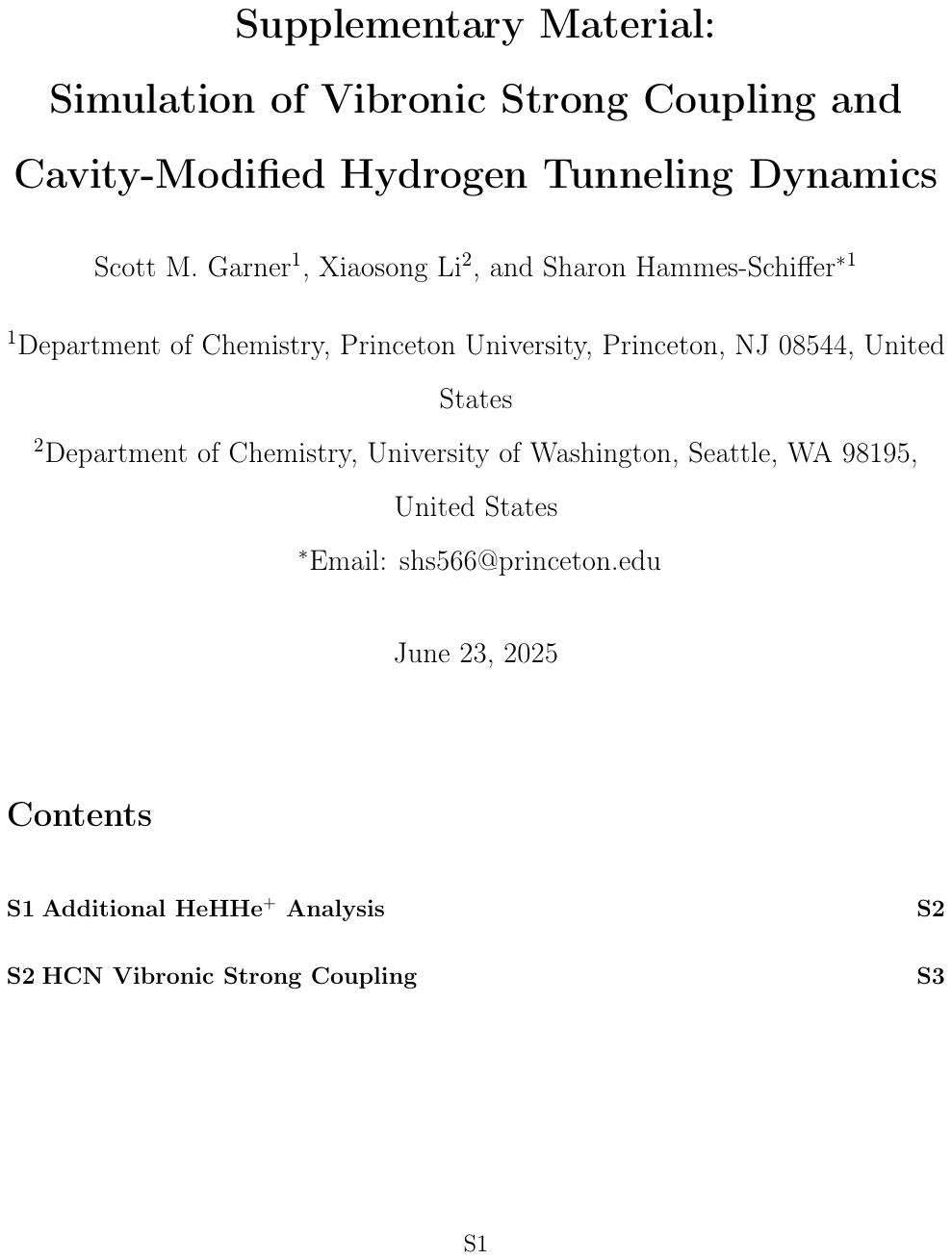}

\end{document}